\documentstyle[11pt,aaspp4]{article}

\def \half {{\textstyle\frac{1}{2}}}

\def \kms {{\rm km~s^{-1}}}
\def \kpc {{\rm kpc}}

\def \Msun {{{\rm M}_{\sun}}}

\def \em {{{\rm e}^-}}
\def \H {{{\rm H}}}
\def \Hp {{{\rm H}^+}}
\def \Hm {{{\rm H}^-}}
\def \HH {{{\rm H}_2}}
\def \HHp {{{\rm H}_2^+}}
\def \He {{{\rm He}}}
\def \Hep {{{\rm He}^+}}
\def \Hepp {{{\rm He}^{++}}}
\def \ne {{n_\em}}
\def \nH {{n_\H}}
\def \nHp {{n_\Hp}}
\def \nHm {{n_\Hm}}
\def \nHH {{n_\HH}}
\def \nHHp {{n_\HHp}}
\def \nHe {{n_\He}}
\def \nHep {{n_\Hep}}
\def \nHepp {{n_\Hepp}}
\def \nd {{\dot{n}}}
\def \upper {{\H}}
\def \low {{\HH}}
\def \DM {{{\rm DM}}}
\def \IGM {{{\rm IGM}}}
\def \gas {{{\rm gas}}}
\def \Hubble {{{\rm Hubble}}}
\def \Jeans {{{\rm Jeans}}}
\def \Bondi {{{\rm Bondi}}}
\def \halo {{{\rm halo}}}
\def \core {{{\rm core}}}

\def \eg {{e{.}g{.}}, }

\def \BDM {\begin{displaymath}}
\def \EDM {\end{displaymath}}
\def \BEQ {\begin{equation}}
\def \EEQ {\end{equation}}
\def \BEQA {\begin{eqnarray}}
\def \EEQA {\end{eqnarray}}
\def \NN {\nonumber}
\def \BL {\begin{list}}
\def \EL {\end{list}}
\def \BENUM {\begin{enumerate}}
\def \EENUM {\end{enumerate}}
\def \BITEM {\begin{itemize}}
\def \EITEM {\end{itemize}}
\def \BARR {\begin{array}}
\def \EARR {\end{array}}
\def \BFIG {\begin{figure}}
\def \EFIG {\end{figure}}

\lefthead{KEPNER, BABUL, \& SPERGEL}
\righthead{Delayed Formation of Dwarf Galaxies}

\begin{document}

\title{The Delayed Formation of Dwarf Galaxies}
\author{
  Jeremy~V.~Kepner\altaffilmark{1,2},
  Arif~Babul\altaffilmark{3}, and 
  David~N.~Spergel\altaffilmark{1}
}
\altaffiltext{1}{Princeton University Observatory, Peyton Hall, Ivy Lane, 
Princeton, NJ 08544--1001 \\
(jvkepner/dns)@astro.princeton.edu}
\altaffiltext{2}{DoE Computational Science Fellow}
\altaffiltext{3}{Dept. of Physics, NYU, New York, NY}

\begin{abstract}

  One of the largest uncertainties in understanding the effect of a
background UV field on galaxy formation is the intensity and evolution
of the radiation field with redshift. This work attempts to shed light
on this issue by computing the quasi-hydrostatic equilibrium states of
gas in spherically symmetric dark matter halos (roughly corresponding to
dwarf galaxies) as a function of the amplitude of the background UV
field.   We integrate the full equations of radiative transfer, heating,
cooling and non-equilibrium chemistry for nine species: H, H$^+$,
H$^-$,H$_2$, H$_2^+$, He, He$^+$, He$^{++}$, and e$^-$.  As the
amplitude of the UV background is decreased the gas in the core of the
dwarf goes through three stages characterized by the predominance
of ionized (H$^+$), neutral (H) and molecular (H$_2$) hydrogen. 
Characterizing the gas state of a dwarf galaxy with the  radiation field
allows us to estimate its behavior for a variety of models of the 
background UV flux. Our results indicate that a typical radiation field
can easily delay the collapse of gas in halos corresponding to
1-$\sigma$ CDM perturbations with circular velocities less than $30~\kms$.

\end{abstract}



\section{Introduction}

  How do galaxies form?  Most models predict that galaxies are assembled
through a successive series of mergers of smaller systems, a  process
known as hierarchical clustering.  These models also predict that small
galaxies form before big galaxies.  Observations, however, suggest the
opposite:  while low mass galaxies are forming the bulk of their stars
at $z\sim 1$ (\cite{Driver95}; \cite{Babul96}), large galaxies appear to
be  well-established by $z \sim 1$ and do not show significant evolution
between $z \sim 1$ and today in luminosity, color, size or space density
(\cite{Steidel94a}; \cite{Steidel94b}).

  The resolution of this apparent conflict between theory and
observations  may lie in the physics of galaxy formation: the
hierarchical clustering paradigm only describes the formation of
gravitationally bound entities, not the process of converting gas into
stars.   Observations of Ly$\alpha$ clouds along quasar lines-of-sight
suggest that at high redshifts, the universe is permeated by a
metagalactic UV flux that may suppress star formation by rapidly
dissociating atomic and molecular gas.  At $z=2$, this background is
estimated to have a strength $J_\nu\approx 10^{-21}$~erg
s$^{-1}$~cm$^{-2}$~ster$^{-1}$~Hz$^{-1}$ at the Lyman limit
(\cite{Bechtold87}). As discussed by \cite{Babul92} (see also 
\cite{Efstathiou92}), an ionizing flux of such intensity can easily
prevent the gas in low mass halos from settling and forming stars.  
Analytic calculations (\cite{Rees86}; \cite{Ikeuchi86}) suggest that the
halos affected will be those with circular velocities $v_c < 30~\kms$, 
a result confirmed by recent numerical studies (\cite{Katz96}; 
\cite{Navarro96a}; \cite{Thoul96}).

  Although the epoch-dependence of the UV flux is uncertain, it is 
expected that $J_\nu$ declines with redshift.  A declining $J_\nu$ has
two effects on the gravitationally bound, photoionized gas in the 
halos.  First, the rate at which the gas is photoheated will decrease
and  hence, its equilibrium temperature will also decrease slightly. 
The reduced temperature  results in a gradual concentration of the gas
towards the center  (\cite{Ikeuchi89}).  Second, the gas becomes more
neutral and offers a greater  optical depth to the ionizing photons
leading to a diminution of  the ionizing flux reaching the central
regions and the formation of a warm ($T\approx 9000$ K) shielded neutral
core (\cite{Murakami90}). The formation of a neutral core, however, is
not a sufficient condition  for star formation.  For a gas cloud to be
susceptible to star formation,  it must be at least marginally
self-gravitating (\eg \cite{Matthews72})  and this can occur only if the
gas in halos with $v_c < 30~\kms$ can cool  below  $9000$ K.

  Further cooling of a neutral metal-poor gas requires the formation and
survival of molecular hydrogen. The formation of molecular hydrogen in a
gas of primordial  composition occurs via gas phase reactions.  Various
studies have shown that an external photoionizing radiation field can
greatly affect the  efficiency of formation of H$_2$ molecules
(\cite{Shapiro87};  \cite{Haiman96a}; \cite{Haiman96b}), with even a
moderate UV flux being capable of suppressing the  H$_2$ abundance.   In
time, however, the decline in the UV intensity will result in the 
formation of sufficient molecular hydrogen to allow the gas in the halos
to cool and become susceptible to star formation.  Babul \& Rees (1992)
argued for $z=1$ as the epoch of galaxy formation in low mass halos, and
sought to identify these forming galaxies with the numerous small faint
blue galaxies seen in deep images.

  In this paper, we present a more thorough investigation of the epoch
of galaxy formation in smaller halos with circular velocities in the
range $15~\kms < v_c < 70~\kms$.  Specifically, we only
consider halos  that form after the metagalactic UV flux has been
established ($z<5$).  Through a detailed study of the thermal and
ionization structure --- we take into account radiation transfer, as
well as heating, cooling and  the corresponding non-equilibrium
chemistry (\cite{Abel96}; \cite{Anninos96})  for nine species: H, H$^+$,
H$^-$, H$_2$, H$_2^+$, He, He$^+$, He$^{++}$, and e$^-$ --- of the
hydrostatic-equilibrium configuration of halo gas subject  to a range of
UV intensities, we determine the threshold UV intensity at  which
molecular hydrogen begins to form.  Given a specific model for the 
evolution of the UV flux, the threshold intensity can be easily
converted  into a redshift of galaxy formation.

 In \S2 of this paper, we describe the dwarf galaxy model, which includes:
the dark matter halos, the amount of gas expected in these halos, the
details of the radiative transfer, heating, cooling and non-equilibrium
chemistry, and the numerical method. \S3 Discusses the results of our
simulations. \S4 Gives our conclusions.

\section{Dwarf Galaxy Model}

  The current numerical models can not yet incorporate the wide range of
physical processes and physical scales associated with galaxy formation.
We focus on the  quasi-hydrostatic evolution of a gas cloud in a small
dark matter halo.  By simplifying the problem to the time evolution of a
spherically symmetric cloud in a dark matter halo, we were able to
incorporate the non-equilibrium chemistry and treat the   radiative
transfer in much greater detail.

\subsection{Dark Matter Halo}
  We will simulate the evolution of the gas in a fixed halo potential. 
For our purposes, the  dark matter halo is specified by two parameters:
the  circular velocity $v_c$ and the virialization redshift $z_v$, which
can be  translated into a halo radius $r_\halo$ and halo mass $M_\halo$ if we
assume that the  overdensity at virialization is $\delta \sim 175$
(\cite{Gunn72})
  \BEQ
       v_c^2 = \frac{G M_\halo}{r_\halo} ~ , ~~~ 
      \frac{4\pi}{3} r_\halo^3 \delta \rho_c(z) = M_\halo ,
  \EEQ 
where the mean density is given by the usual expressions for a
$\Omega=1$ CDM cosmological model: $\rho_c(z) = (1
+ z)^3 \rho_c^0$, $6 \pi G \rho_c t_\Hubble^2 = 1$, $t_\Hubble^0 = 2/3 H_0$.  

  We will focus on the evolution of ``typical'' halos, which correspond
to 1-$\sigma$ perturbations.   In a $b = 1.6$ standard CDM model,
Press-Schechter theory implies that the halos virialized at a
redshift given by
  \BEQ
       1 + z_v \approx (7.5/1.7)(M_\halo/10^9~\Msun)^{0.1} ~ , ~~~
         10^8~\Msun < M_\halo < 10^{10}~\Msun
  \EEQ
Figure 1 shows the relationship between halo circular velocity and the
virialization epoch for the 1-$\sigma$ perturbations.  The basic trend is
generic to all hierarchical models: small objects form first and larger
objects form later.

We use a dark matter halo profile that has been fit by \cite{Burkert95}
to galaxy rotation curves,
  \BEQ
     \rho_\DM(r) = \frac{r_0}{(1 + x) (1 + x^2)} ~, ~~~  x = r/r_0
  \EEQ
which in turn can be related to the halo radius and mass by
  \BEQA
       r_\halo &=& 3.4 r_0 , \NN \\
       M_\halo &=& M_\DM(r_\halo) , \NN \\
       M_\DM(r)  &=& \int_0^{r} \rho_\DM(r) 4 \pi r^2 dr .
  \EEQA
While recent numerical work suggests that the halo density profiles  of
large galaxies are proportional to $r^{-1}$ in the centers and $r^{-3}$
at the edges (\cite{Navarro96b}), these profiles do not fit the dwarf galaxy
observations (\cite{Moore94}; \cite{Flores94}).  

\subsection{Gas Content and Profile}

  The ultraviolet background is able to heat the gas to a temperature of
roughly $10^4~\arcdeg$K.   In large halos, where $v_c > 50~\kms$, the
gas pressure is relatively unimportant and the gas content is determined
by the global value of $\Omega_b$: $M_\gas = \Omega_b M_\halo$ (assuming
$\Omega=1$). However, for smaller halos collapsing out of a hot IGM the
gas pressure  resists the collapse (\cite{Thoul96}) and $M_\gas <
\Omega_b M_\halo$.  We now make some simple estimates as to where this
transition occurs and how much gas should reside in the halo.  If the
gas in the uncollapsed halo is greater than the Jeans mass than the gas
should collapse of its own accord.  This provides an upper limit to
amount of gas in the halo
 \BEQ
      M_\gas = \Omega_b M_\halo ~, ~~~ \Omega_b M_\halo > M_\Jeans
 \EEQ
where 
$M_\Jeans = f \Omega_b rho_c (2 \pi c_\IGM t_\Hubble)^3$, $f = 3/\pi \sqrt{2}$,
$c_\IGM^2 = 1.5 k_B T_\IGM / \mu$.  For $\Omega_b M_\halo < M_\Jeans$ we can
formulate an upper limit for $M_\gas$ by estimating the amount of mass
that could be acreted via Bondi accretion in a Hubble time.  Thus,
 \BEQ
      M_\gas = M_\Bondi ~, ~~~ \Omega_b M_\halo < M_J
 \EEQ
where $M_\Bondi = f t_\Hubble \dot{M}_\Bondi$, $\dot{M}_\Bondi = \pi G^2
M_\halo^2 \rho_c / c_\IGM^3$. Note, the $O(1)$ factor $f$ has been included
in $M_\Jeans$ and $M_\Bondi$ so that $M_\Jeans = M_\Bondi$ when
$\Omega_b M_\halo = M_\Jeans$.  Plots of these various masses as function
of redshift are shown in Figure 2 for $T_\IGM = 20,000~\arcdeg$K,
indicating that the halos become more DM dominated as they get smaller,
which is consistent with the observations
(\cite{Carignan88}; \cite{deBlok96}).

  The initial gas density profile is specified by hydrostatic equilibrium
and by our assumption that the gas is in thermal equilibrium with the
background radiation field. This approximation will hold as long as the
gas has time to react to changes in the DM potential, self-gravity, and
cooling. The scales governing these processes are the sound speed,
Hubble time, Jeans mass, and the net cooling/heating time. The sound
speed of a $10^4~^\circ$K gas is $\sim 15~\kms$, corresponding to a
crossing time for a typical $r_\halo \sim 10~\kpc$ halo of $\sim
6\times10^8$ years.  The DM potential will change on the order of a
Hubble time, which at $z \sim 3$ is  $\sim 2\times10^9$ years. The gas will
be stable to self-gravity if it is less than the Jeans mass, which from
Figure~2 is true for nearly the entire range of models. The heating/cooling
time becomes important only when a significant amount of $\HH$ has
formed, and determining this point one of the goals of this paper.

\subsection{Non-Equilibrium Chemistry}

  The important role of the detailed chemistry of primordial gas (in
particular the formation of $\HH$) has been known and studied since it
was first proposed as a mechanism for the formation of globular clusters
(\cite{Peebles68}). The potential number of reactions in this simple
mixture of H and He is enormous (\cite{Janev87}).  \cite{Abel96} have
selected a subset of these reactions to model the behavior of
primordial gas for low densities ($n < 10^4$) over a range of
temperatures ($1^\circ$K $< T < 10^8~^\circ$K); these equations
(see Appendix A) represent a careful balance between computational
efficiency and accuracy.

\subsection{Radiative Transfer}

  Fully 3D radiative transfer requires estimating the contribution to
the flux at every point from every other point along all paths for each
wavelength.  At the minimum this is a 6D problem.  However, in most
instances symmetries can be introduced which result in a more tractable
situation. The simplest situation occurs when the gas can be assumed to
be optically thin throughout.  This approximation is sufficient in the
majority of cosmological situations (\cite{Katz96}; \cite{Navarro96a};
\cite{Anninos96}) and only breaks down in the cores of halos that have
undergone sufficient cooling, a situation that is usually made further
intractable by the complexities of star formation. The next simplest
geometry is that of a slab (or a sphere under the assumption of a
radially perpendicular radiation field), which leaves an intrinsically
2D problem.  This approach is the most common in radiative transfer and
has been used to address similar situations (\cite{Haiman96a};
\cite{Haiman96c}).  Although this approach may not be a bad
approximation for a sphere in an isotropic radiation field, we choose to
account for all the different paths that penetrate a given spherical shell,
which leaves an inherently 3D problem.  Although this geometry can
significantly increase the  size of the computations, for a static grid
(i.e. non-adaptive) pre-computing of the various geometric factors can
significant alleviate this situation (see Appendix B).  Taking into
account the different paths effectively ``softens'' the optical depth,
smoothing out transitions from optically thin to optically thick
regimes.  In addition, introducing the accounting for the different
paths lays the groundwork for exploring more general geometries.

\subsection{Heating and Cooling}
  Perhaps the most important aspect of the model is the balance between
the heating and cooling processes.  This balance is what allows the
establishment of a quasi-static temperature profile for a specific
radiative flux.  If the balance between the heating and cooling is not
established, then the hydrostatic equilibrium solution to the gas profile
will evolve too rapidly.  Fortunately, this situation only comes about
when the gas in the halo becomes dense and a large amount H$_2$ is
formed.  As was mentioned earlier, this point presumably marks the
onset of star formation, which is one of the central goals of this work.

  The temperature profile is evolved via the heating and cooling
functions found in \cite{Anninos96}
\BEQ
      \frac{\dot{T}}{T} = \frac{\Gamma(T) - \dot{E}(T)}
                     {E{T}}
\EEQ
where $E = 1.5 k_B T \sum n_i$, $i = \em, \H, \Hp, \Hm, \HH, \HHp, \He,
\Hep, \Hepp$. $\Gamma$ includes  photoionizaton heating and $\dot{E}$
includes cooling due to collisional excitation, collisional ionization,
recombination, molecular hydrogen, bremsstrahlung and Compton cooling. 
All the appropriate functions are taken from Appendix B of \cite{Anninos96}.

\subsection{Numerical Method}
  The microphysical processes couple to the larger scale density profile
primarily through radiative heating, which sets the temperature profile.
The rate of radiative heating is in turn strongly dependent on the
column densities of each species, which is set by the temperature.  Thus,
we have a system which is described by differential equations on the
small scale with integral constraints on the large scale.  The
difficulty of solving such a system in the large variety of time scales
involved. Solving the entire set simultaneously is prohibitive.  Our
approach has been to use code modules which solve for each of the
processes  independently.  Iterating between the modules then provides
adequate approximation to the true solution. The species solver and
heating/cooling modules were provided by Yu Zhang and Mike Norman of
NCSA (\cite{Abel96}; \cite{Anninos96}).  The radiative transfer and
hydrostatic equilibrium modules we wrote ourselves.  Our central goal is
to see how the gas behaves as a function of the amplitude of the
background UV field.  The overall approach is to pick a halo, set an
initial value of $J_0$, and estimate the time and number of steps to
integrate at this value.  We then use each module at each time step
which converges on the overall solution.  Each module was tested and
verified independently.  In addition, where possible, combinations of the
modules have been tested.  A more detailed description
of our numerical method is given in Appendix C.

\section{Results and Discussion}

   The goal of the simulation is to determine the maximum
ionizing flux that permitted the  formation of molecular gas (and
stars).  The simulations began with the gas ionized. Over
the range of objects, the state of the core exhibits the same
qualitative behavior (see Figure 3). As the flux decreases the core goes
through three phases ionized H$^+$ ($T_\core \sim 20,000^\circ$K),
neutral H ($T_\core \sim 10,000^\circ$K) and  molecular H$_2$
($T_\core \sim 100^\circ$K).

  In the initial  H$^+$ state, the object is completely ionized and
resembles the IGM.  In the H state the core is neutral and the object is
like an inverse Stromgren sphere, except that the ionization is
primarily collisional, a result of the high temperature maintained by
the radiative heating.  In the H$_2$ state, conditions allow for the
formation of molecular hydrogen in the core; consequently, the cooling
time becomes much less than the dynamical time and the core collapses.

  Five simulations were conducted for halos in the range $15~\kms < v_c
< 70~\kms$ corresponding to: $3\times10^8~\Msun < M_\halo < 5 \times
10^{10}~\Msun$, $4 > z_v > 2$ , and $3\times10^5~\Msun < M_\gas < 3
\times 10^9~\Msun$ (see Figure~2).  For each object there are fluxes
$J_{\upper,\low}$ at which the upper transition  H$^+$ $\rightarrow$ H
and the lower transition H $\rightarrow$ H$_2$ occur (see Figure 4). 
$J_{\upper,\low}(z_v)$ can be fit by a formula
  \BEQ
       \ln[J_{\upper,\low}/J_{21}]  =  a_{\upper,\low} - b_{\upper,\low} z_v ~,
  \EEQ
where $a_{\upper,\low} = (17.9,17.3)$, and $b_{\upper,\low} = (6.2,6.4)$

  Knowing $J_{\upper,\low}$ allows us to place bounds on the behavior of an object in
a given radiation field.  Most significantly, if the radiation field is
above  $J_\upper$, then it will be impossible for the object to collapse and
form stars.  If  the radiation field is below $J_\low$, then it must
collapse.  Most likely the  actual value of the flux at which collapse
occurs is between $J_\upper$ and $J_\low$.

  For objects consistent with 1-$\sigma$ perturbations in a standard CDM
cosmology selecting the evolution of the radiation field ($J(z,\nu)$)
specifies critical redshifts $z_{\upper,\low}$  such that for
$z_\low < z < z_\upper$ the object will be in the neutral H state. 
Figures 5, 6 and 7 plot the virial redshift and the critical redshifts
as a function of the circular velocity for three different amplitudes of
an evolving UV field
  \BEQ
       J(z,\nu) = \left\{
                 \begin{array}{c@{\quad,\quad}c}
                   J_0 ~ [(1+z)/4]^4 ~ (\nu_{Ly}/\nu) &  z < 3 \\
                   J_0 ~ (\nu_{Ly}/\nu) & z > 3
                 \end{array} \right. ~.
  \EEQ
These results indicate that a typical radiation field can easily
prevent gas  in halos with $v_c < 30~\kms$ from collapsing.  Furthermore,
the less massive  halos, which are typically the first to virialize, are
the last to form galaxies.  Thus, if these results have any correspondence
to observed dwarf galaxies, they suggest that the larger objects will
burst earlier, and perhaps will show an observable trend of increasing
age with mass.

  Another important aspect of these objects is their potential
contribution to Quasar absorption lines.  Figure 8 shows the column
density evolution for H and $\Hep$.  During the $\Hp$ phase $N_\H \sim
10^{16-17}$, suggesting that these objects may make a contribution to
the Ly$\alpha$ forest and Ly limit systems.  During the $\H$ phase
$N_\H$ increases dramatically to levels near those seen in damped
Ly$\alpha$ clouds, ($N_\H \sim 10^{21}$) but this is probably a
relatively short lived phase and therefore is it unlikely that these
objects would be seen in Quasar spectra.

  Our calculations ignored the role of metals in the evolution of the
gas. Metals can have a number of important effects on the chemistry and
dynamics of gas clouds: dust grains will absorb ionizing radiation and
serve as formation sites for molecular hydrogen and atomic lines of C
and other heavy elements can be important coolants at temperatures
around $10^4~^\circ$K. Are these processes important in dwarf galaxies at high
redshift? Observations of QSO Lyman forest clouds suggest that the metal
abundances in meta-galactic gas was $Z \sim 0.001 - 0.01$ times the
solar value at $z \sim 3$  (\cite{Songaila96}).  At these abundances,
heavy element cooling is unimportant (\cite{Bohringer89}). If we make the
conservative assumption that most of the carbon at high redshift is
incorporated into dust grains (as in our galaxy), with a size
distribution that is similar to the local ISM, then this suggests a
cross-section per hydrogen atom of $\sigma_{\rm dust}(1000\AA) \approx
Z~2 \times 10^{-21}~{\rm cm}^2$ (\cite{Draine96}). In our models,
the maximum column density occurs when the cloud is most centrally
condensed, which occurs just before the onset of $\HH$ formation and is
roughly $N_\H \sim 10^{21}~{\rm cm}^{-2}$ (see Figure~8).  Thus the
maximum optical depth at these wavelengths is approximately $\tau_{\rm
dust} \sim Z \sim 0.01$. The contribution of dust to $\HH$ formation in
our galaxy can be approximated by a $\nd_\HH \approx R n \nH$, where $R
= 6 \times 10^{-18} T^{1/2}~{\rm cm}^3~{\rm s}^{-1}$ (\cite{Draine96}). 
If we scale $R$ by the metallicity, then the dust term will be
negligible in comparison to the other terms contributing to $\HH$
formation whenever $\nHm > 10^{-9}$, which is nearly always the case in
the neutral H core.

\section{Conclusions}

  In this paper, we have studied the evolution of a proto-dwarf galaxy
exposed to the metagalactic radiation field. We begin with an ionized
gas cloud, initially Jeans stable, in a dark matter halo and follow its
hydrostatic evolution as the meta-galactic flux decreases.  Our
calculations indicate that the state of the gas can be characterized by
the predominance of ionized, neutral, or molecular hydrogen in the
core.  The transitions between these phases takes place quickly as the
amplitude of the background flux decreases.  We have computed the
critical fluxes at which the transitions take place $J_{\upper,\low}$,
which serves as upper and lower bounds on the flux at which rapid
cooling and the subsequent star formation can begin. If the flux is
greater than $J_{\upper}$, then is unlikely that the gas will cool. If
the flux is less that $J_{\low}$, then the gas must cool.

  Characterizing the state of the gas in terms of the background flux
allows us to use any model for the evolution of flux with redshift.
Using typical values of the flux, our simulations indicate that gas in a
$v_c = 30~\kms$ halo collapsing at $z \approx 3$ can easily be prevented
from forming significant amounts of $\HH$ until $z \approx 2$.
Thus our calculations are consistent with photoionization delaying
the formation of low mass galaxies (\cite{Babul92}).

  In this paper we have focused on 1-$\sigma$ perturabtions, in
subsequent work we will expand our exploration of the ($v_c$,$z_v$)
plane, which will allow us to further address the issues pertaining to
the observed faint blue galaxies.  As an example, our preliminary
calculations indicate that a $v_c = 30~\kms$ halo which virializes at
$z_v = 2$ will first be able to form neutral H at $z \approx 1.25$.

\acknowledgments

We gratefully acknowledge Yu Zhang and Mike Norman for the use of their
non-equilibrium species solver.  J. Kepner was supported by the Dept. of
Energy Computational Science Fellowship Program. DNS is supported by the
MAP/MIDEX program. This work was partially supported by NASA through
Hubble Archival Research Grant AR-06337.20-94A awarded by Space
Telescope Science Institute, which is operated by the Association of
Universities for Research in Astronomy, Inc., under NASA contract
NAS5-26555.

\appendix
\section{Non-equilibrium Chemistry}
The evolution of the electron, hydrogen and helium abundances
are computed from the following set of coupled
equations (\cite{Abel96}):

  \BEQA
      \nd_\H &=&
                k_2 \ne \nHp
             - (k_1 \ne +  k_{20}) \nH
             \\
      \nd_\Hp &=& - \nd_\H
             \\
      \nd_\He &=&
                k_4 \ne \nHep
             - (k_3 \ne + k_{21}) \nHe
             \\
      \nd_\Hep &=& - \nd_\He - \nd_\Hepp
             \\
      \nd_\Hepp &=&
                k_5 \ne \nHe
             -  k_6 \ne \nHepp
             + k_{22} \nHep
             \\
      \nHm &=& \frac
                {k_7 \ne \nH}
                {(k_8 + k_{15}) \nH + (k_{16} + k_{17}) \nHp + k_{14} \ne + k_{23}}
             \\
      \nd_\HH &=&
                k_8 \nHm \nH + k_{10} \nHHp \nH + k_{19} \nHHp \nHm
      \NN \\ 
                &-& (k_{13} \nH + k_{11} \nHp + k_{12} \ne + k_{24} + k_{27}) \nHH
             \\
      \nHHp &=& \frac
                {k_9 \nH \nHp + k_{11} \nHH \nHp  + k_{17} \nHm \nHp + (k_{24} + k_{27}) \nHH}
                {k_{10} \nH + k_{18} \ne + k_{19} \nHm + k_{25} + k_{26}}
             \\
      \nd_\em &=&
                (k_1 \nH - k_2 \nHp + k_3 \nHe - k_4 \nHep + k_5 \nHep - k_6 \nHepp) \ne
       \\
               &+& k_{20} \nH + k_{22} \nHep + k_{21} \nHe
  \EEQA
The equilibrium values of $\nHm$ and $\nHHp$ are used as the timescales
are relatively short.   To conserve mass and charge the
following local constraints are also imposed
  \BEQA
        \nH + \nHp + \nHm + 2 \nHH + 2 \nHHp &=& {\rm constant} \NN \\
        \nHe + \nHep + \nHepp  &=& {\rm constant} \NN \\
        \nHp - \nHm + \nHHp + \nHep + 2 \nHepp   &=& \ne
  \EEQA
The rate coefficients $k_{1-19}(T)$ are taken from Tables A.1 and A.2
of \cite{Abel96}, and the photoionization and photodisocciaton coefficients
$k_{20-26}$ are computed from the radiation field (see next section).

\section{Radiative Transfer}
  The photoionization and photodisocciation $k_{20-26}$ coefficients are
computed from the radiation field via
  \BEQ
        k_i = 4 \pi \int_{\nu_i}^\infty 
              \frac{\sigma_i(\nu) J(\nu,r)}{h \nu} d\nu ~, ~~~ i=20,\ldots,26
  \EEQ
where $\sigma_{20-26}(\nu)$ are taken from Table A.3 of \cite{Abel96}.
The mean flux at a distance $r$ from the center is computed from
  \BEQ
       J(\nu,r) = \half J(\nu,r_\halo) \int_{-1}^{+1} \exp[-\tau(\nu,r,\mu)] d\mu,
  \EEQ
where
  \BEQ
       \tau(\nu,r,\mu) = \sum_i \sigma_i(\nu) N_i(r,\mu) ~, ~~~ i = 20,\ldots,26 ~,
  \EEQ
and the cross-section indices and the column densities map as follows
$\{20, 21, 22, 23, 24, 25, 26\} \rightarrow
\{\H, \Hep, \He, \Hm, \HHp, \HHp, \HH\}$.
The column density of each species along each line of sight is
  \BEQ
       N_i(r,\mu) = \int_0^{l(r,r_\halo,\mu)} n_i(r'(x)) dx ~,
  \EEQ
where $l(r,r_\halo,\mu) = r \mu + \sqrt{r_\halo^2 - r^2 (1 - \mu^2)}$, 
$r'(x) = \sqrt{r^2 + x^2 - 2 r x \mu}$, and $\mu = \cos \theta$ (see
Figure 9). For the impinging background UV field, we choose
$J(\nu,r_\halo) =  J_0 (\nu/\nu_{Ly})^{-\alpha}$ and  unless otherwise
stated $\alpha = 1$. The final coefficient for the photodissociation of
H$_2$ is taken from \cite{Draine96}, which attempts to include the
effects of self-shielding
  \BEQA
       k_{27}(r) &=& 0.15~\zeta_{pump}~\chi~\bar{f}_{shield}(r)~, \NN \\
       \zeta_{pump} &\approx& 3 \times 10^{-10}~{\rm sec}^{-1} ~,\NN \\
       \chi &=& \frac{[J(\nu,r_\halo) \nu / c]_{c/\nu = 1000\AA}}
                     {4 \times 10^{-14}~{\rm erg}~{\rm cm}^{-3}} ~,\NN \\
       \bar{f}_{shield}(r) &=& \half \int_{-1}^{+1} f_{shield}(r,\mu) d\mu ~,\NN \\
       f_{shield}(r,\mu) &=& \left\{
                 \begin{array}{c@{\quad,\quad}c}
                   1 & x < 1 \\
                   x^{-0.75} & x > 1
                 \end{array} \right. ~,  \NN \\
       x &=& \frac{N_\HH(r,\mu)} {10^{14}~{\rm cm}^{-2}} ~ .
  \EEQA

\section{Numerical Algorithm}
  The following is a step-by-step outline of the code:
\BENUM
  \item Initialization.
  \BENUM
     \item Choose $z_v$, compute $M_\halo$ and $r_\halo$ for 1-$\sigma$ perturbations.
     \item Compute $M_\gas$ based on $M_\Jeans$ and $M_\Bondi$.
     \item Compute $\rho_\DM(r)$ using Burkert profile.
     \item Choose $J_0$.  Initialize $T(r) \sim 20,000~\arcdeg$K.
     \item Choose initial mass fractions for H, H$^+$, H$^-$,H$_2$,
           H$_2^+$, H$_2^-$, He, He$^+$, He$^++$, and e$^-$.
  \EENUM
  \item Compute number of time steps to integrate over this value of $J_0$
        based on heating/cooling and chemical time scales.
  \item Compute hydrostatic equilibrium. Once the total amount of gas in
the halo is specified the hydrostatic equilbribum gas profile can be
computed by setting the gas temperature profile $T(r)$.  If we assume that
the potential is DM dominated throughout, then
 \BEQ
      \frac{\rho_\gas(r) T(r)}{\rho_\gas(0) T(0)} 
         = \exp \left[ - \int_0^r \frac{\mu}{k_B T(r')}
                                  \frac{G M_\DM(r')}{r'^2} \right] ,
 \EEQ
where $\rho_\gas(0)$ is determined from the normalization
$M_\gas = \int_0^{r_\halo} 4 \pi r'^2 \rho_\gas(r')dr'$.

  \BENUM
     \item Compute $\rho_\gas(r)/\rho_\gas(0)$ so that it is in
           hydrostatic equilibrium with  $T(r)$ and the DM potential
           set by $\rho_\DM(r)$.
     \item Use $M_\gas$ to compute $\rho_\gas(0)$.
     \item At each radius recompute the number densities of each species 
           while preserving relative fractions.
  \EENUM
  \item Radiative transfer.
  \BENUM
     \item Compute optical depth at each radius along each line of sight.
     \item Solve for the internal radiation field $J(\nu,r)$.
     \item Compute photoionization and photodissociation rates.
  \EENUM
  \item Non-equilibrium chemistry.
  \BENUM
     \item Interpolate temperature dependent rate coefficients.
     \item Integrate non-equilibrium reaction equations.
  \EENUM
  \item Heating and cooling.
  \BENUM
     \item Compute heating/cooling due to photoionization, 
           collisional ionization, recombination, collisional
           excitation, bremsstrahlung, Compton and H$_2$ cooling.
     \item Integrate and solve for $T(r)$.
  \EENUM
  \item If done with all the time steps for this value of $J_0$, 
        decrement $J_0$ and go to step 2.
\EENUM
The result of this method is the a solution which represents the
quasi-hydrostatic equilibrium state of the gas, which in the absence of
hydrodynamic heating processes can be thought of as the asymptotic
solution of the spherical collapse of gas into a static potential.

  At the heart of our code are the various arrays use to store the
relavent data.  The gas dynamics consists of the arrays to
store the DM, gas and temperature profiles: $\rho_\DM(r)$,
$\rho_\gas(r)$, $T(r)$.  All the radially dependent arrays are
evaluated on grid $r_j$ with $N_r$ points. The chemistry data consist of
one number density array, $n_i(r)$, for each species.

  The radiative transfer data is composed of several arrays, beginning
with the array 
  \BEQ
       \Delta_{jj'k} = l(r_j,r_{j'+\half},\mu_k) - l(r_j,r_{j'-\half},\mu_k)
  \EEQ
which contains the path length spent in a shell $r_{j'}$ of a ray
traveling to a shell centered on $r_j$ at an angle $\mu_k$ (see Figure
9).  The ability to pre-compute these path lengths is critical to the
efficient computation of the column density array for each species
along each path at each radius $N(r,\mu)$.  Using the column density
and cross-section arrays $\sigma_{20-26}(\nu)$ the flux array can be
computed $J(\nu,r)$ and subsequently the $k_{20-26}$. One additional
complication arises from the sharp photoionization boundaries in the
cross sections, which require very high values of $N_\nu$ to resolve
accurately. However, this problem can be solved by placing extra high
resolution grid points at these boundaries.  We found $7$ additional
points at each boundary provided sufficient accuracy.

  The convergence properties of all the various grid parameters were
tested, as well as the time step parameters.  The values we selected
were $N_r = 200$, $N_\mu = 20$, $N_\nu = 200$, which were twice the
minimum necessary to provide accurate results.  The $\nu$ grid was based
on a logarithmic scale with $0.74~{\rm eV} < h \nu < 0.74 \times
10^4~{\rm eV}$; increasing the range of $\nu$ did not alter the results.


\newpage


\begin{figure}
\plotone{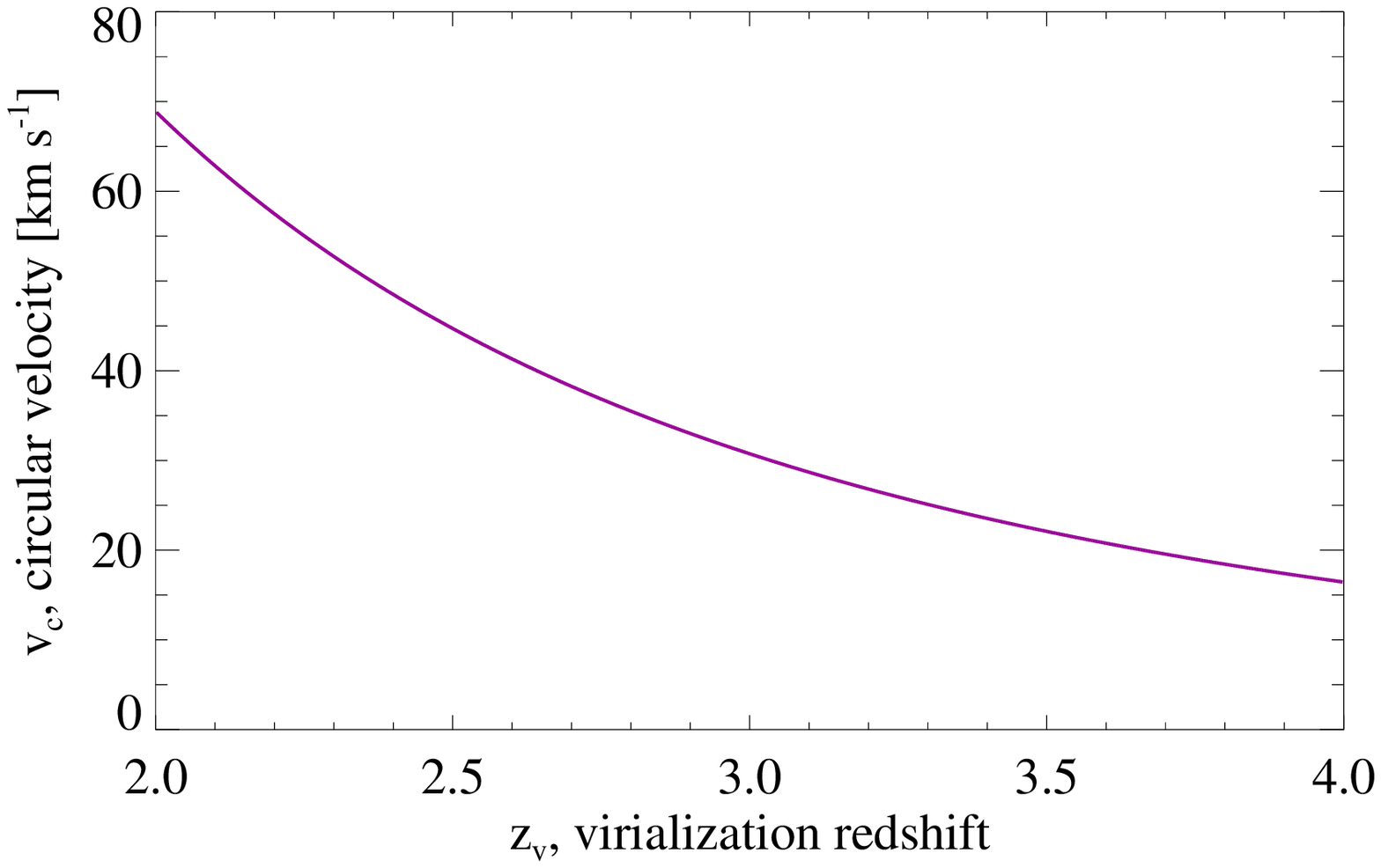}
\caption{
  Circular velocity as a function of virialization redshift for
1-$\sigma$ perturbations as computed from Press-Schecter theory
using standard CDM.
}
\end{figure}

\begin{figure}
\plotone{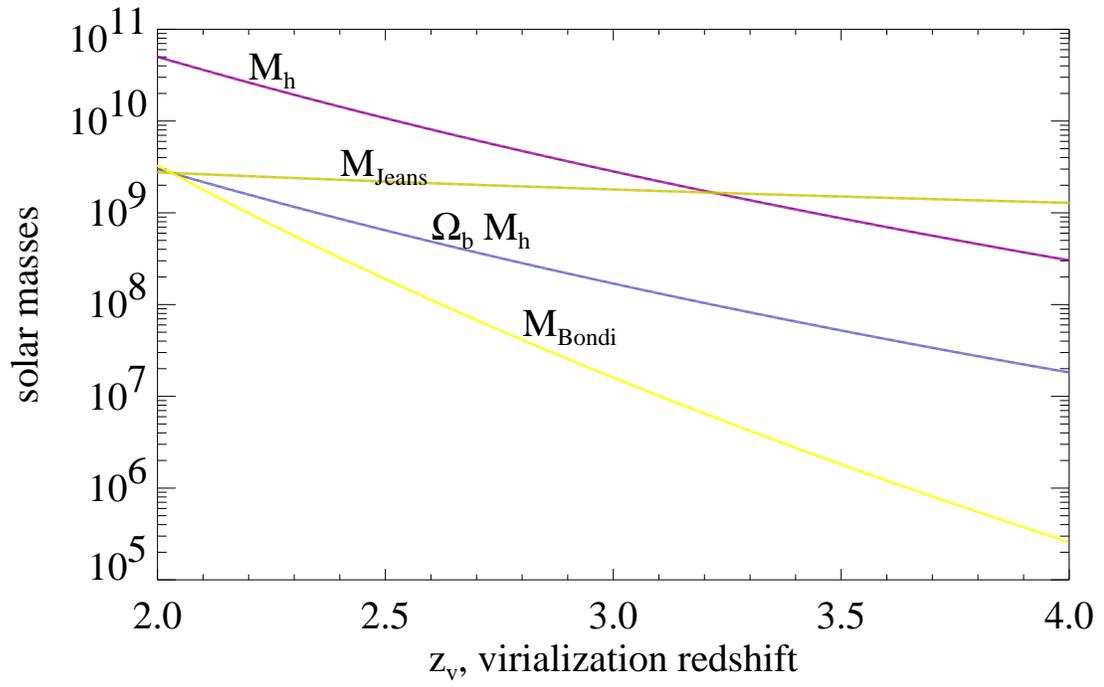}
\caption{
  $M_\halo$, $\Omega_b M_\halo$, $M_\Jeans$, and $M_\Bondi$
for 1-$\sigma$ perturbations collapsing out of a $20,000~\arcdeg$K IGM.
}
\end{figure}

\begin{figure}
\plotone{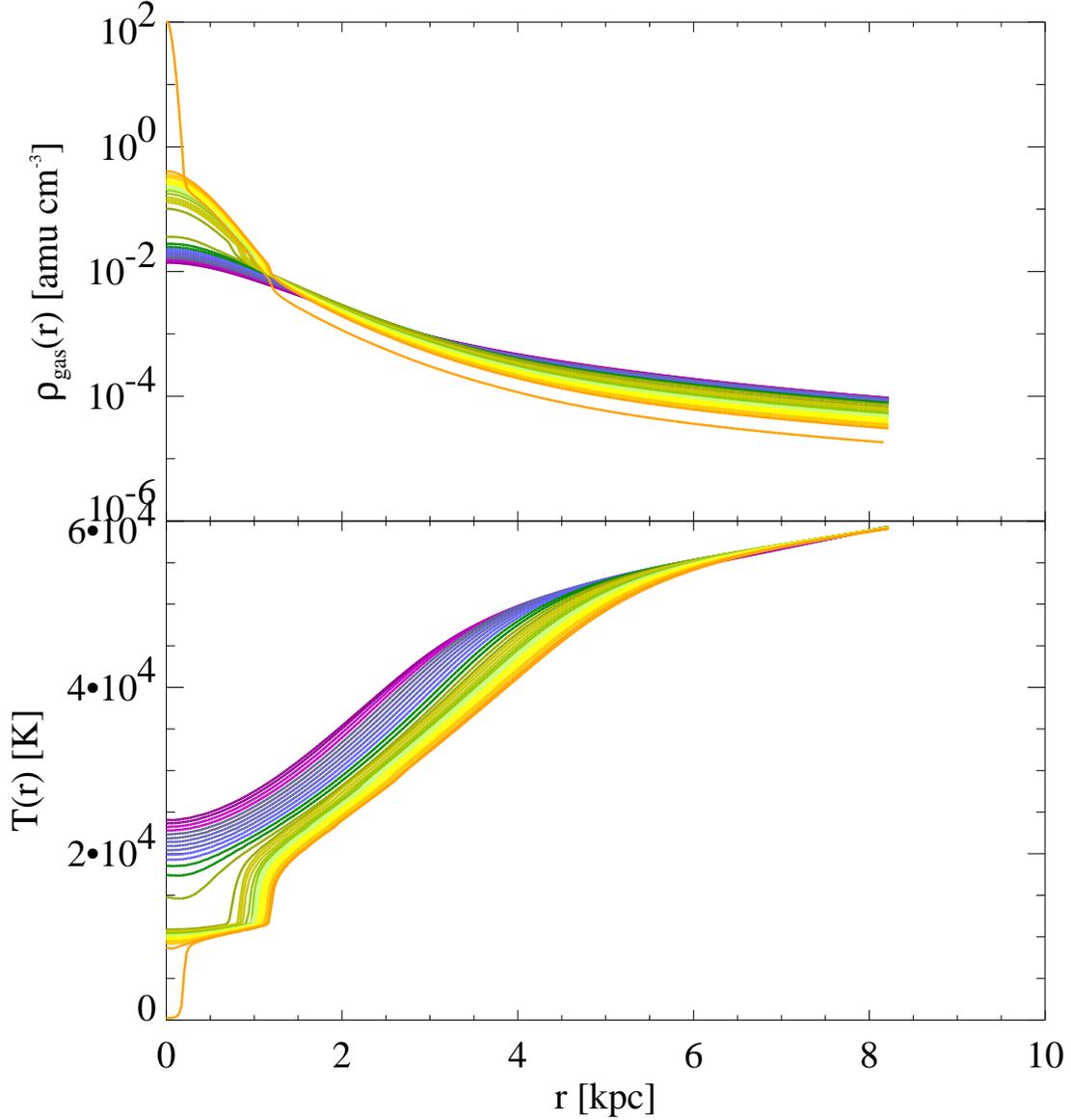}
\caption{
  Evolution of the gas density profile (top panel) and gas  temperature
profile (bottom panel) of a typical object ($v_c \approx 30~kms$, $z_v 
= 3$, $M_\halo = 2.8\times 10^9~\Msun$, $M_\gas = 1.5\times10^7~\Msun$)
from higher flux (purple = 1.0~$J_{21}$) to lower flux (red =
0.14~$J_{21}$), where  $J_{21} =
10^{-21}$~erg~s$^{-1}$~cm$^{-2}$~ster$^{-1}$~Hz$^-1$. As the flux
decreases the behavior of the core is characterized by  three phases:
$\Hp$, H and $\HH$.   In the $\Hp$ phase the flux is high enough to 
heat the core to $T_\core \sim 20,000~\arcdeg$K.  In the H phase the
flux does not  penetrate to the center and a distinct H core develops
with $T_\core \sim 10,000~\arcdeg$K.  In the $\HH$ phase the flux is
sufficiently weak that self-shielding  allows $\HH$ to form, which
rapidly increases the cooling rate and causes the gas in the core to
collapse. The transitions between phases occurs abruptly at $J_\upper \approx
0.46~J_{21}$ and $J_\low \approx 0.14~J_{21}$.
}
\end{figure}

\begin{figure}
\plotone{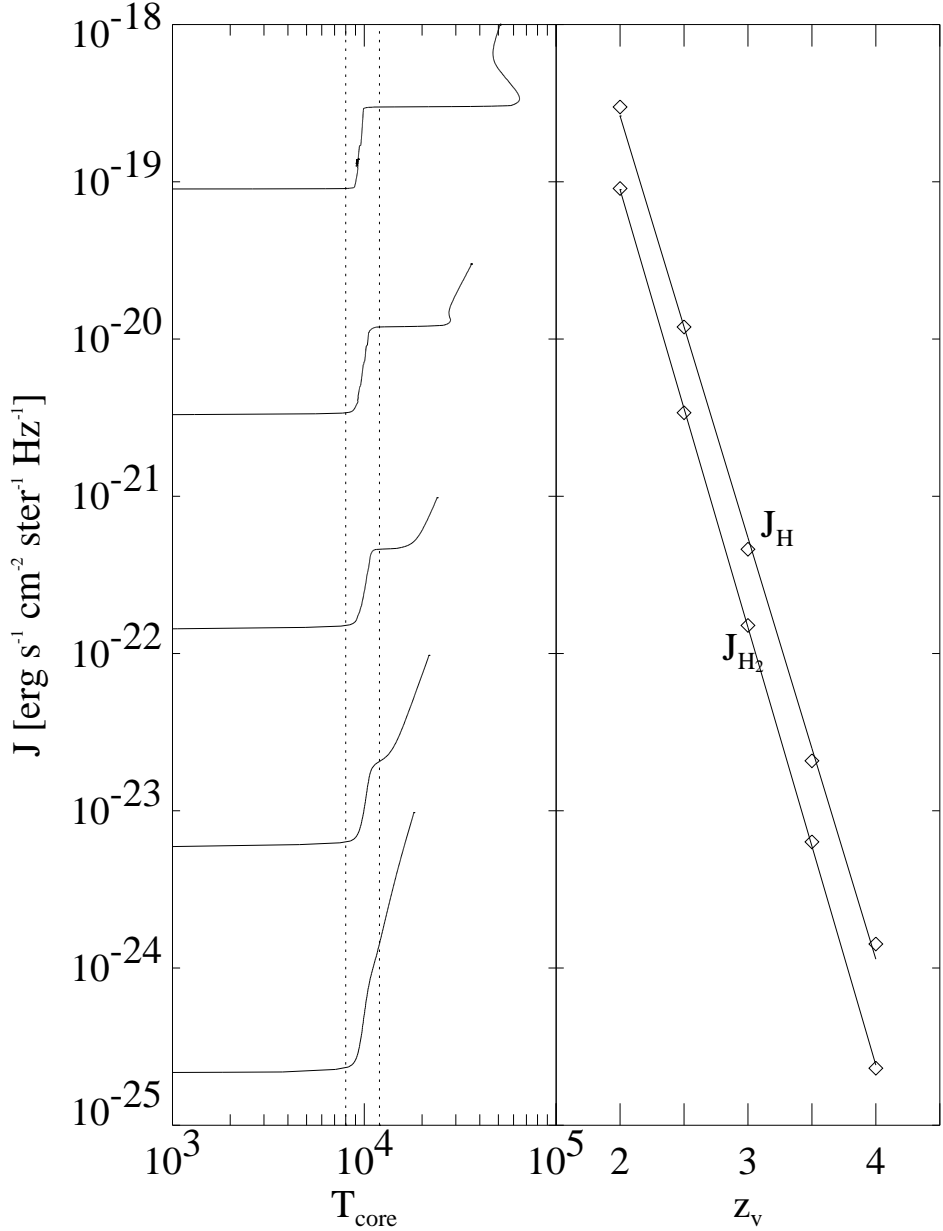}
\caption{
  For each object there are critical values of the radiation flux  which
mark the upper phase transition from $\Hp$ $\rightarrow$ H ($J_\upper$) and
the lower  phase transition from H $\rightarrow$ $\HH$ ($J_\low$). The left
panel shows the flux as a function of the central temperature for
several different objects ($z_v$ = 4.0, 3.5, 3.0, 2.5, 2.0). We define
$J_\upper$ to be the point where $T_\core$ drops below  12,000~$\arcdeg$K
(right dotted line).  Likewise, $J_\low$ is the point where $T_\core$
drops  below 8,000~$\arcdeg$K (left dotted line).  The left panel
shows $J_{\upper,\low}$ as function of the virial redshift.  The straight
lines are the best fits: $\ln[J_{\upper,\low}/J_{21}]  =  a_{\upper,\low} - b_{\upper,\low} z_v$,
where $a_{\upper,\low} = (17.9,17.3)$, and $b_{\upper,\low} = (6.2,6.4)$
}
\end{figure}

\begin{figure}
\plotone{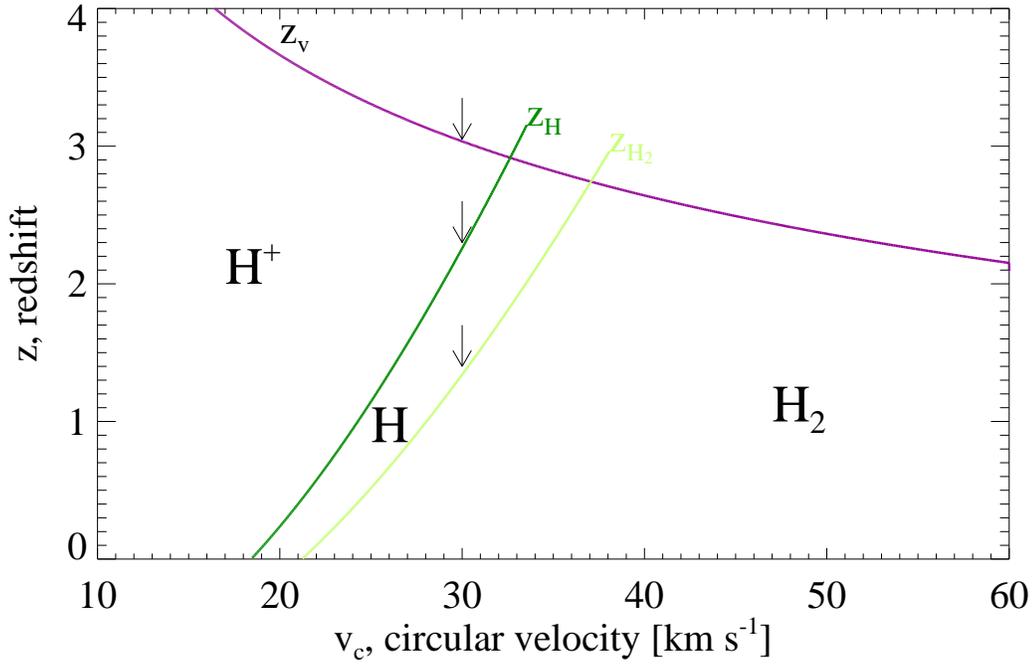}
\caption{
This figure shows the collapse redshift, $z_v$, the redshift of H
formation, $z_\upper$, and the redshift of $\HH$ formation, $z_\low$, as
a function of the circular velocity of $v_c$.  Thus, a 1-$\sigma$
perturbation with $v_c = 30~\kms$ collapses at $z \approx 3$ (top
arrow), but is ionized until $z \approx 2.5$ (middle arrow) and cannot
form molecular hydrogen until $z \approx 1.5$ (bottom arrow). The
ionizing field evolves as $J(z) =  J_0 [(1+z)/4]^4$, for $z < 3$ and
$J(z) = J_0$, for $z > 3$ with the spectral index held fixed at $\alpha
= 1$.  This figure is for a flux amplitude of $J_0 = 10^{-21}$.
}
\end{figure}

\begin{figure}
\plotone{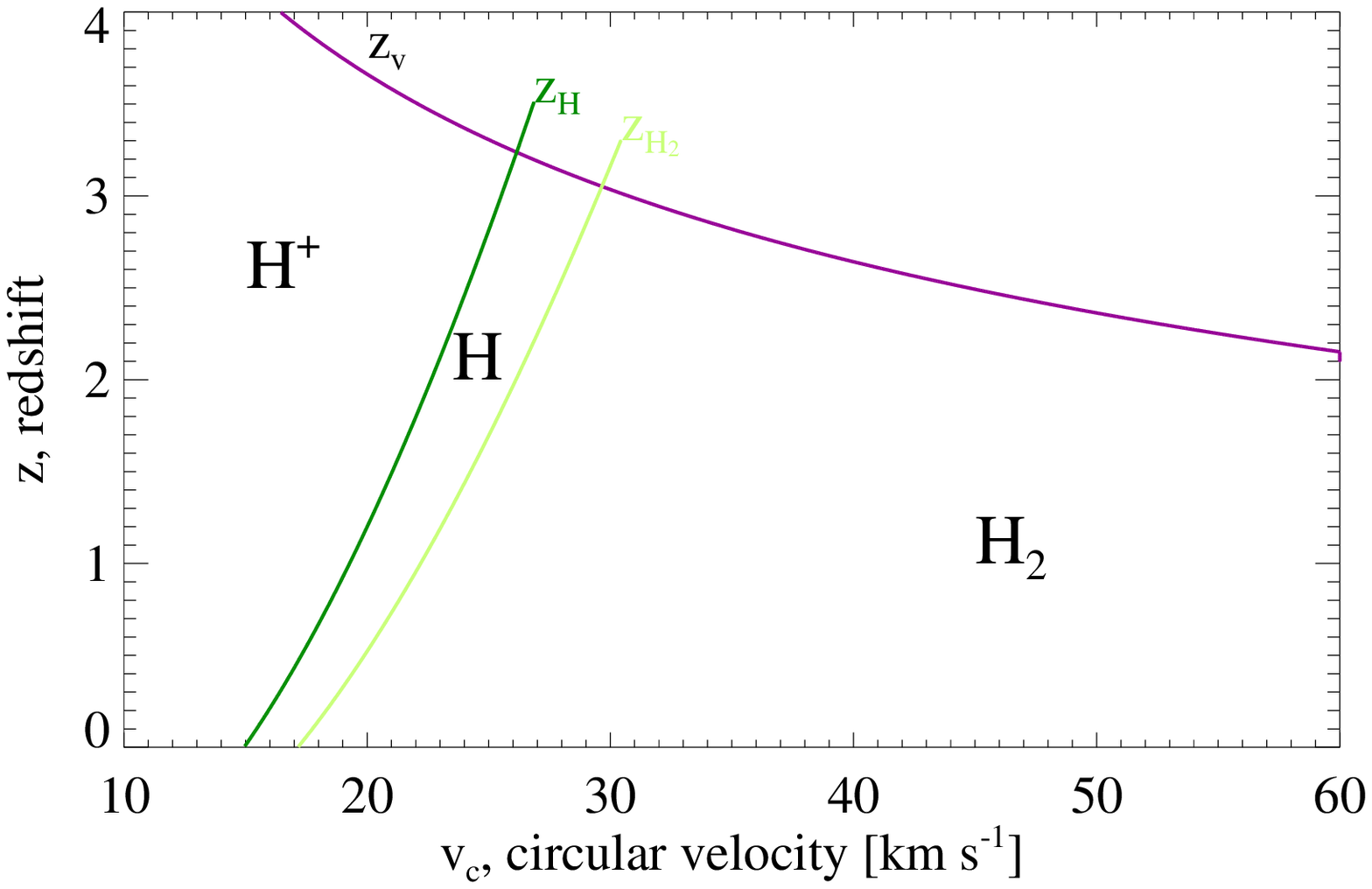}
\caption{
Same as previous figure, but with $J_0 = 10^{-22}$.
}
\end{figure}

\begin{figure}
\plotone{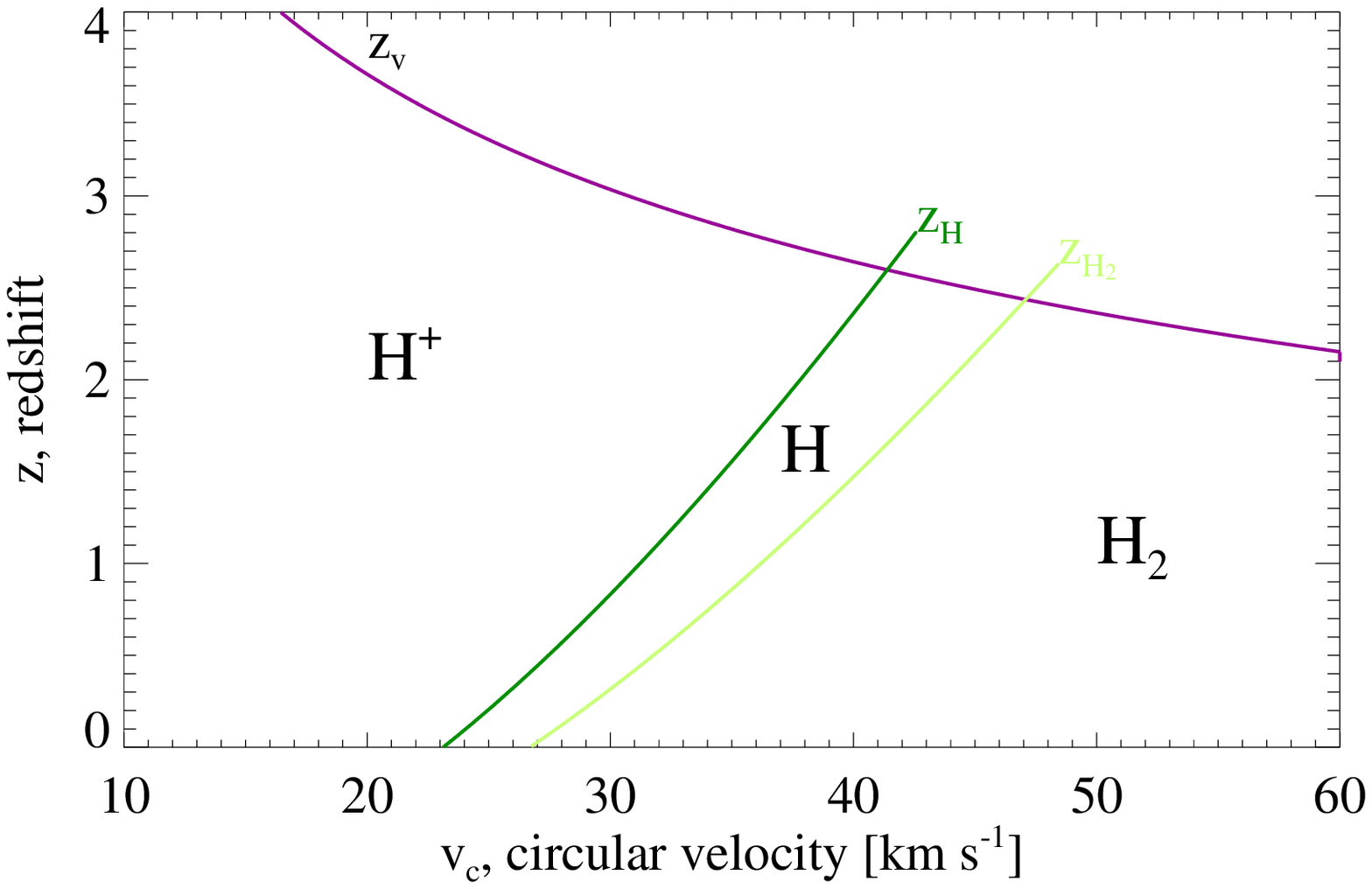}
\caption{
Same as previous figure, but with $J_0 = 10^{-20}$.
}
\end{figure}

\begin{figure}
\plotone{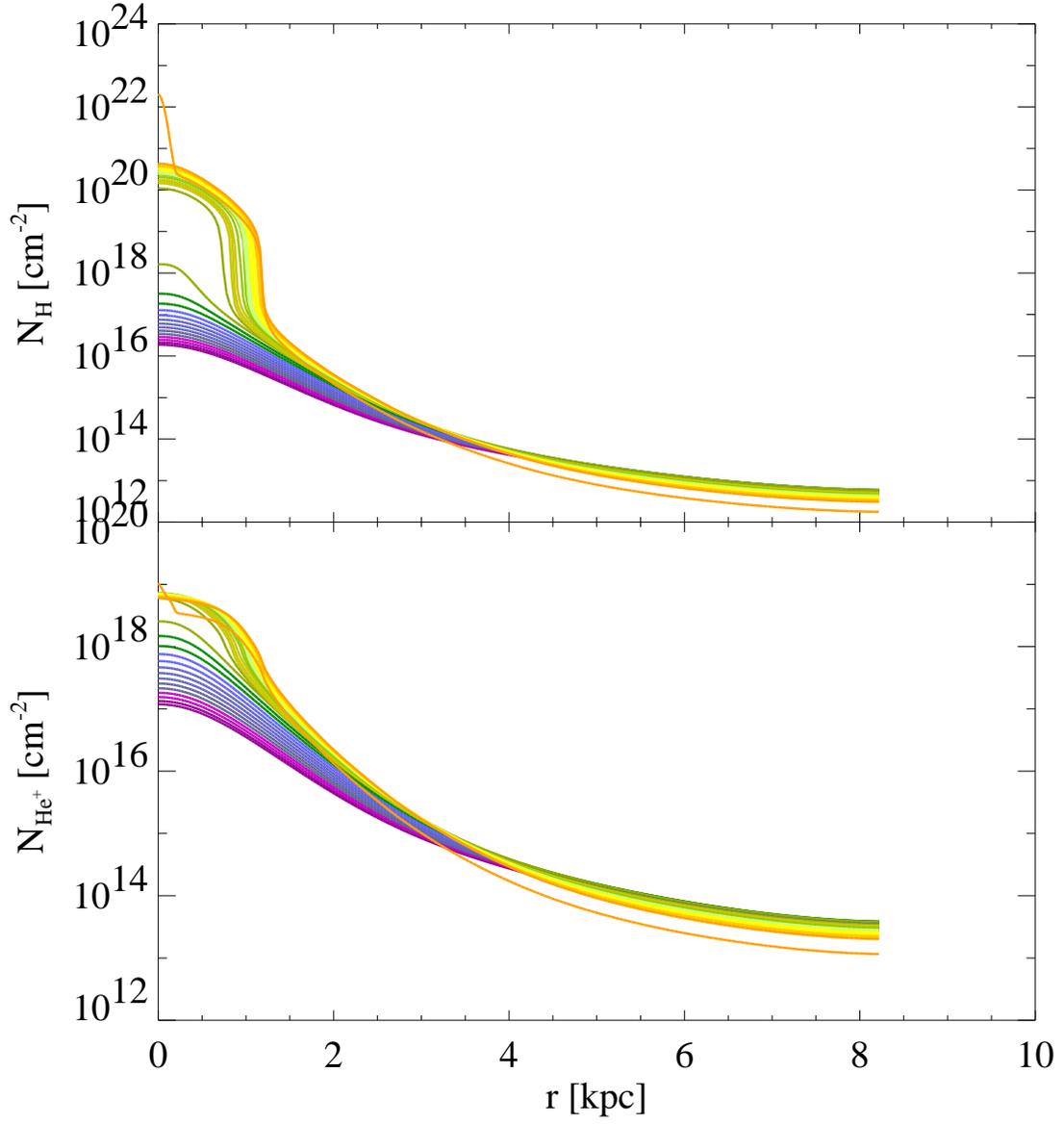}
\caption{
  Evolution of the column density profiles of H (top panel) and
He$^+$ (bottom panel) of the same object shown in Figure 3.
}
\end{figure}

\begin{figure}
\plotone{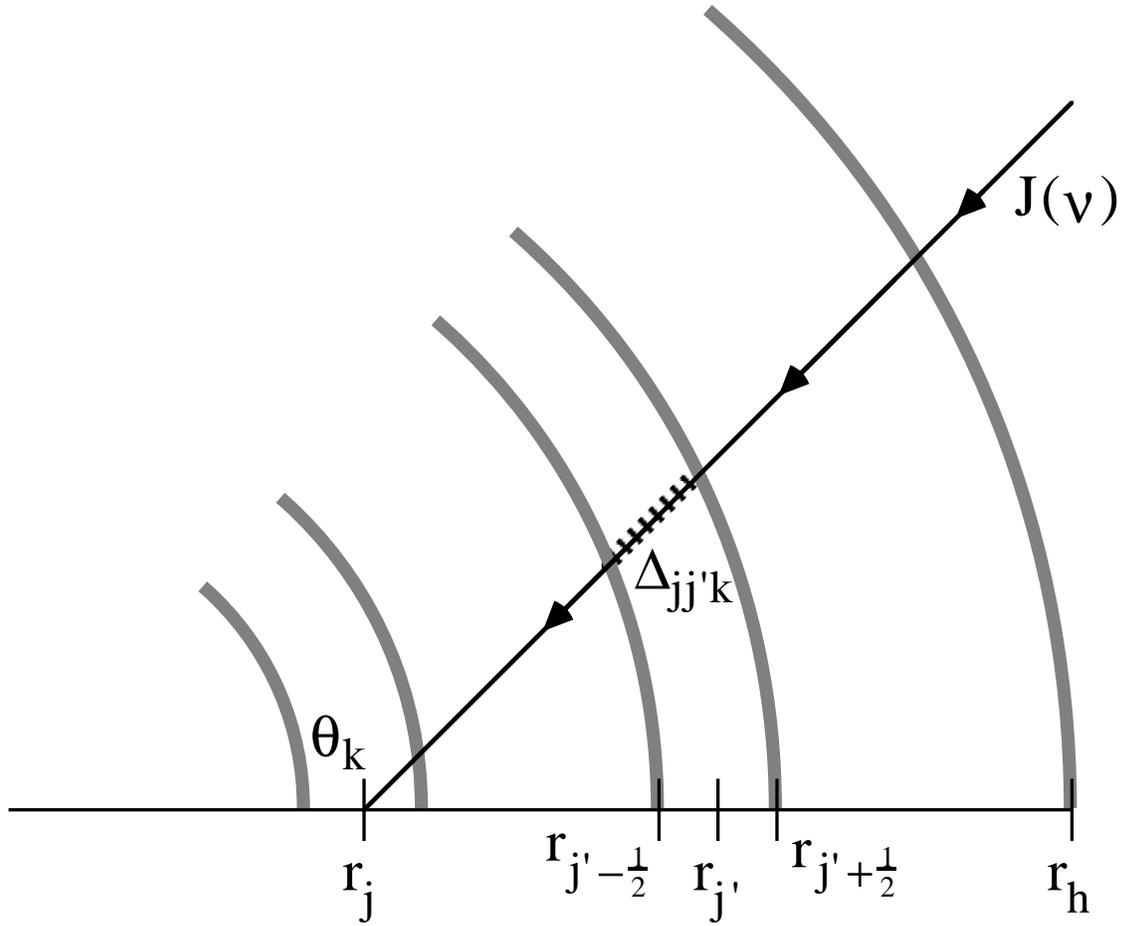}
\caption{
  Radiative transfer geometry for an impinging isotropic flux $J(\nu)$.  
$\Delta_{jj'k}$ is the distance traveled through shell $r_{j'}$
by an incoming ray bound for shell $r_j$ at an angle $\mu_k = \cos \theta_k$.
}
\end{figure}

\end{document}